# Title: Berry curvature memory through electrically driven stacking transitions

Jun Xiao[1,2,9], Ying Wang[3,9], Hua Wang[4], C. D. Pemmaraju[2], Siqi Wang[3], Philipp Muscher[1], Edbert J. Sie[2,5], Clara M. Nyby[6], Thomas P. Devereaux[1,2,5], Xiaofeng Qian[4], Xiang Zhang[3,8*] & Aaron M. Lindenberg[1,2,7*]

**Affiliations:**

[1] Department of Materials Science and Engineering, Stanford University, Stanford, CA 94305, USA.

[2] SIMES, SLAC National Accelerator Laboratory, Menlo Park, CA 94025, USA.

[3] Nanoscale Science and Engineering Center (NSEC), University of California at Berkeley, Berkeley, CA 94720, USA.

[4] Department of Materials Science and Engineering, Texas A&M University, College Station, TX 77843, USA.

[5] Geballe Laboratory for Advanced Materials, Stanford University, Stanford, CA 94305, USA.

[6] Department of Chemistry, Stanford University, Stanford, CA, USA.

[7] PULSE Institute, SLAC National Accelerator Laboratory, Menlo Park, CA, 94025, USA.

[8] Faculties of Science and Engineering, The University of Hong Kong, Hong Kong, China.

[9] These authors contributed equally: Jun Xiao, Ying Wang

*Correspondence to: aaronl@stanford.edu; xiang@berkeley.edu

**Abstract:** In two-dimensional layered quantum materials, the stacking order of the layers determines both the crystalline symmetry and electronic properties such as the Berry curvature, topology and electron correlation[1–4]. Electrical stimuli can influence quasiparticle interactions and the free-energy landscape[5,6], making it possible to dynamically modify the stacking order and reveal hidden structures that host different quantum properties. Here we demonstrate electrically driven stacking transitions that can be applied to design nonvolatile memory based on Berry curvature in few-layer $WTe_2$. The interplay of out-of-plane electric fields and electrostatic doping controls in-plane interlayer sliding and creates multiple polar and centrosymmetric stacking orders. In situ nonlinear Hall transport reveals such stacking rearrangements result in a layer-parity-selective Berry curvature memory in momentum space, where the sign reversal of the Berry curvature and its dipole only occurs in odd-layer crystals. Our findings open an avenue towards exploring coupling between topology, electron correlations, and ferroelectricity in hidden stacking orders and demonstrate a new low-energy-cost, electrically controlled topological memory in the atomically thin limit.

## Main

Berry curvature is a fundamental concept in condensed-matter physics quantifying the topological local entanglement between conduction and valence bands in crystalline solids without time-reversal or space-inversion symmetry[7]. This non-zero geometrical property of the band structure determines the anomalous motion of coherent electron wave packets and is a prerequisite for global topological invariants in many quantum materials such as Chern-class topological materials. Recently, the study of Berry curvature in layered van der Waals materials has attracted great attention, leading to many intriguing discoveries such as non-trivial topology in Weyl semimetals[8], valleytronics in bulk 3R $MoS_2$ and topological transport in bilayer graphene domain walls[3,9]. These findings indicate the fundamental role of layer stacking order, the relative crystallographic arrangement of atomic layers, in determining crystal symmetry and its enforced topological electronic properties. Recent findings on electron correlation and superconductivity in twisted bilayer graphene[1,10], further motivate the control of phase competition to access hidden stacking orders for novel physics and applications. Here we discover multiple stacking order transitions driven by electrical fields and doping in few-layer $WTe_2$. The transitions among these phases through interlayer sliding enable synchronized annihilation and switching of the ferroelectric polarization and the Berry curvature. We further observe layer-parity-selective memory behavior of Berry curvature in momentum space, where the sign reversal of Berry curvature only occurs in odd-layer crystals, consistent with a recent theoretical prediction[11].

$T_d$ phase $WTe_2$ is a layered material that crystallizes in a distorted hexagonal net with an orthorhombic unit cell (Fig. 1a). Prior theoretical studies have indicated that the $WTe_2$ can deviate from this equilibrium polar crystal structure to form different hidden stacking orders with distinct symmetries[12,13], including a non-polar monoclinic 1T' structure, and a polar orthorhombic $T_d$ phase with upward spontaneous polarization ($T_{d,}\uparrow$) or downward spontaneous polarization ($T_{d,}\downarrow$). These three phases have same single-layer atomic configurations but distinct relative interlayer sliding along the *b* crystalline axis (Fig. 1a). The energy barrier for different stackings can be as small as few meV/unit cell[14], about one to two orders of magnitude smaller than that for conventional bond rearrangement in phase-change materials[15,16]. Therefore, electrical perturbation of ground state ordering is nontrivial and may drive proximal interlayer arrangements with unexplored physics and quantum properties. Further, $T_d$ $WTe_2$ exhibits many novel topological and quantum geometrical physics: for its monolayer form, it shows quantum spin Hall effect, tunable superconductivity and canted spin texture[17–23]; while for its bulk crystal, it is a type-II topological Weyl semimetal with chiral Berry curvature monopoles[2,8]. The arrangement of these positive and negative Berry curvature hotspots in momentum space leads to a nontrivial Berry curvature dipole (Fig. 1a), defined as $D_{ij} = -\int_k \Omega_j \partial_i f_0$ ( $\Omega_j$ is the Berry curvature along the $j$ direction and $f_0$ is the equilibrium electron distribution; the integral is over the $k$-space or momentum space)[24]. Even down to the ultrathin limit, it maintains large Berry curvature and the associated dipole near the Fermi level[25,26]. Therefore, $WTe_2$ is an ideal platform to demonstrate dynamic stacking transitions by electrical stimulus and its deterministic influence on Berry curvature and its dipole.

To realize the electrical manipulation, we fabricated dual-gate devices based on few-layer $WTe_2$ encapsulated by two boron nitride (h-BN) flakes with thicknesses of $d_t$ and $d_b$ (Fig. 1b). The applied electric field passing upwards is $E_\perp = (-V_t/d_t + V_b/d_b)/2$, while the external hole doping is $n_h = -\varepsilon_{hBN}\varepsilon_0 (V_t/d_t + V_b/d_b)/e$. Here $V_t$, $V_b$, $\varepsilon_{hBN}$, $\varepsilon_0$ are the top gate bias, the bottom gate bias, dielectric constant of h-BN and vacuum permittivity, respectively. Two types of phase transitions with distinct hysteresis shapes are observed under different gating configurations: under pure doping condition ($V_t/d_t = V_b/d_b$) in a five-layer sample, electrical conductance G shows a rectangular-shape hysteresis (labeled as Type I), appearing on the hole doping regime (Fig. 1c). In contrast, when only $E_\perp$ is swept ($-V_t/d_t = V_b/d_b$), the electrical conductance shows a butterfly-shape response (labeled as Type II) and a bistability near zero bias (Fig. 1d).

To experimentally reveal the origin of the above two distinct phase transitions, we use *in-situ* gate-dependent second harmonic generation (SHG) and Raman spectroscopy. SHG is a sensitive probe to lattice asymmetry and space group in layered materials[15,27]. A pristine $T_d$ five-layer $WTe_2$ exhibits both strong SHG and the expected polarization pattern for P$m$ space group (Supplementary Fig. 1). We measure the SHG variation correlated with Type I electrical conductance under pure external doping sweeping (Fig. 2a). Accompanying an abrupt electrical conductance change at hole doping ~ $1.5\times 10^{13}/cm^2$, the SHG intensity shows a sharp decrease by a factor of five leaving a residue comparable to the background SHG from h-BN and graphite (Supplementary Fig. 1c). This indicates the formation of a centrosymmetric phase prohibiting SHG, with measured threshold doping level consistent with prior theoretical predictions for a transition into 1T' stacking[12]. The metastable nature of this induced phase results in a constant low SHG intensity during the retraction of carrier doping. At a lower hole doping level of $0.4 \times 10^{13}/cm^2$, we observe a rapid increase back to the initial intensity level of the $T_d$ phase. We also find that a single gate bias providing both electrostatic doping and an electric field can trigger Type I hysteresis in both the electrical conductance and SHG but with a lower hole doping threshold (Supplementary Fig. 2). This indicates the same phase transition origin with similar hysteresis shapes observed under pure doping and single bias gating configurations. The additional E-field in single gate bias, when antiparallel to the initial $T_d$ polarization, helps to lower the doping threshold requirement (Fig. 1c, Fig. 2a and Supplementary Fig. 2). We further observe the SHG modulation depth during the transition has strong layer dependence under a single gate bias (Fig. 2b). The interference contrast that arises from different thicknesses of *h*-BN may result in variations of the initial SHG intensity for different devices. However, such contrast is expected to be independent of gate bias. Here we normalize the initial intensity and focus on the relative intensity variation when sweeping the electrical bias. Compared to their initial SHG intensity, the SHG drop is much more notable in trilayer and five-layer samples compared with that in four-layer samples. Such layer dependence is in line with the layer dependent crystalline symmetry of 1T' stacking[28], i.e. the space group of even-layer 1T' $WTe_2$ is $P_m$ without inversion symmetry leading to nontrivial SHG intensity and hence small SHG modulation. On the other hand, odd-layer 1T' $WTe_2$ belongs to the $P2_1/m$ space group with inversion symmetry resulting in the large observed SHG intensity change. Meanwhile, we find negligible change of linear absorption of $WTe_2$ at the pump wavelength (Supplementary Fig. 3 and 4).

Besides the observed symmetry transformation, we confirm such phase change is a stacking-order transition from $T_d$ to 1T' through interlayer sliding as opposed to an intralayer bond distortion or bond breaking by *in-situ* gate-dependent Raman spectroscopy and coherent phonon dynamics. The inversion symmetry breaking in $T_d$ phase allows a Raman-active interlayer shear mode corresponding to interlayer vibrations along the crystalline *b* axis (~9 $cm^{-1}$ at 80 K). The peak around 11 $cm^{-1}$ is attributed to a breathing mode while the other higher frequency peaks originate from intralayer vibrations within each atomic layer (Fig. 2c and Supplementary Fig. 6) [29]. Unlike the even-parity breathing mode, the odd-parity shear mode is expected to vanish in a centrosymmetric state like 1T' stacking[30]. Indeed, we observe a substantial intensity reduction of the interlayer shear mode in a five-layer sample during the Type I transition (Fig. 2c). Meanwhile, a negligible modulation in other high frequency intralayer vibration modes is observed (Fig.2c and Supplementary Fig. 5). The gate dependent intensity of the shear mode further shows rectangular-shape hysteresis response corresponding to the Type I structural phase transition (Fig. 2d). In complementary coherent phonon dynamics measurements (Fig. 2e), we again observe the disappearance of the shear mode in the time domain after formation of the centrosymmetric phase in a five-layer device. The recovery of inversion symmetry is expected for the well-known 1T' monoclinic stacking as well as a subtle variation recently proposed under nonequilibrium optical pumping[31]. Taken together, the above evidence indicates the Type I phase change origin is a stacking transition between $T_d$ and 1T' through interlayer sliding along the crystalline *b* axis, without any intralayer bond distortion or bond breaking.

Next we uncover the origin of Type II phase transition driven by pure electric field. The corresponding SHG measurement shows butterfly-shape intensity hysteresis in both a four-layer and a five-layer samples (Fig. 3a). This observation reflects the switching between ferroelectric polarizations, similar to that reported in prototypical ferroelectric oxides[32]. This is further supported by transport measurements on a top ultrathin graphite, which reveal the different resistances influenced by the opposite spontaneous polarizations before and after the switching (Supplementary Fig. 7 and 8). In addition, the SHG intensity minima at the turning points in four-layer and five-layer crystals show significant difference in magnitude, similar to the layer dependent SHG contrast in 1T' stacking. This shows that stacking structure changes take place in this ferroelectric switching process, which may involve 1T' stacking as the intermediate transition state *via* interlayer sliding. To identify the stacking order nature of fully poled upward and downward polarization phases in the Type II transition, we studied their characteristic lattice excitations. The two phases show the similar Raman frequency and amplitude of shear mode as well as high frequency vibrations belonging to polar $T_d$ crystal geometry (Fig. 3b). In addition, the corresponding SHG polarization patterns are almost identical in terms of both pattern types and lobe orientations (Fig. 3c). These findings reveal that the Type II phase transition is a ferroelectric stacking switching between $T_{d,\uparrow}$ and $T_{d,\downarrow}$ orders, which also explains the unknown microscopic origin for similar conductance phenomena reported in a recent study[33]. The larger sliding displacement to switch these $T_d$ phases thus leads to the observed larger electric field requirement (~0.4 V/nm) for fully poling, which is about 1.5-2.0 times higher than that applied in the formation of the intermediate stacking.

In the following, we now show how such electrically driven stacking transitions enable layer-parity selective memory behavior of Berry curvature and Berry curvature dipole by nonlinear Hall effect. Since the nonlinear Hall signal is proportional to the Berry curvature dipole strength[24], the nonlinear Hall effect has recently been identified as a hallmark to probe Berry curvature and its distribution in momentum space in time-reversal-invariant materials[25]. To capture the maximum nonlinear Hall response from the intrinsic Berry curvature dipole[26], the geometry of metal contacts is designed to allow current flow along the $a$ axis and generate nonlinear Hall voltage along the $b$ axis (Fig. 4a and Supplementary Fig. 9). Figure 4b shows the expected quadratic power relationship between applied AC current along $a$ axis ($I_{//,\omega}$) and its 2nd harmonic transverse voltage along $b$ axis ($V_{\perp,2\omega}$). The second-harmonic transverse response $V_{\perp,2\omega}$ is on the order of 0.1% of $V_{//,\omega}$, the first-harmonic longitudinal voltage along the $a$ axis. The associated Berry curvature dipole along $a$ axis, $D_{ac} = -\int_k \Omega_c \partial_a f_0$, is proportional to $V_{\perp,2\omega}$ / $(V_{//,\omega})^2$ and estimated to be on the order of ~1 Å (Supplementary Information Section IX), which is consistent with prior reports[26]. Between the $T_{d,\uparrow}$ and $T_{d,\downarrow}$ transitions, the conductance G and nonlinear Hall signal $V_{\perp,2\omega}$ / $(V_{//,\omega})^2$ in both trilayer and four-layer $WTe_2$ shows a clear hysteresis (Fig. 4c and d). Intriguingly, there is a sign switching of the nonlinear Hall signal in the trilayer (Fig. 4c and Supplementary Fig. 11), indicating the proportional in-plane Berry curvature dipole also reverses its direction and possesses binary memory switching property. In contrast, the sign of nonlinear hall hysteresis signal is invariant in the four-layer $WTe_2$ (Fig. 4d). This layer parity dependence of sign switching is further evidenced by first-principles calculations of Berry curvature for trilayer and four-layer $WTe_2$ (Fig. 4f, g and Supplementary Fig. 10).

We discover this striking layer dependent reversal of Berry curvature dipole originating from the layer dependent symmetry operation during the transition from ferroelectric $T_{d,\uparrow}$ to $T_{d,\downarrow}$ states and the pseudovector character of the Berry curvature[11]. Although $T_{d,\uparrow}$ and $T_{d,\downarrow}$ stackings are physically formed by different interlayer sliding, they are effectively related to each other via symmetry operations depending on layer number parity. In particular, the relation is an inversion operation for odd-layer $WTe_2$, while a mirror operation with respect to the *ab* plane and a global half unit-cell translation along the *a* axis for even layers (Fig. 4e). As a consequence, the out-of-plane Berry curvature pseudovector distribution in odd-layer inverts its direction while it maintains the same sign in even-layer for each transition between these two stacking orders. This directly leads to the layer parity selective reversal of Berry curvature dipole and nonlinear Hall signal. Besides the substantial difference in initial and final ferroelectric stackings, the Berry curvatures were found to move in momentum space during such ferroelectric stacking transition (Supplementary Movie 1), and a hysteresis of the nonlinear Hall signal was also observed in type I phase transition showing vanishing nonlinear Hall response at induced 1T' stacking in a trilayer sample (Supplementary Fig. 12). These findings clearly demonstrate the nontrivial evolution of the position and strength of Berry curvature in momentum space through stacking order transitions, enabling dynamic control of Berry curvatures and a memory storage property of the Berry curvature dipole. Such electrical control of Berry curvature based on interlayer-sliding-mediated stacking order transitions is fundamentally different from previous reports on pure electronic band modification in monolayer $WTe_2$ without any stacking orders[34]. This difference results in non-volatile and hysteretic memory behavior of Berry curvature with layer parity selection not observed before. This new type of memory is expected to be highly energy-efficient. Given the small stacking barrier (few meV/unit cell) and the capacitor charging energy, only ~ 0.1 aJ/nm$^2$ energy is estimated to be consumed for each stacking transition as a single "writing" operation (Supplementary Information Section X). On the other hand, although $T_{d,}\uparrow$ and $T_{d,}\downarrow$ semimetal show little electrical conductance difference, the opposite sign of their large Berry curvature dipole in odd-layer allows for substantial contrast in nonlinear Hall reading, enabling a non-destructive and nonmagnetic reading mechanism. The nontrivially large Berry curvature in momentum space and new memory reading mechanism found here may resolve the long-term major challenge of reading binary information in polar metals, which conventionally rely on challenging detection of screened weak spontaneous polarizations in real space. Therefore, our electrical control and reading of Berry curvature memory is promising to make polar metal with ferroelectric polarization not only fundamentally interesting but practically useful.

In summary, we report the first observation of electrically driven stacking transitions in few-layer $WTe_2$ and associated Berry curvature memory determined by layer number parity. Future rational system design and lateral scaling are needed to quantify the performance advantages (e.g. energy cost per bit, operation speed, and density) of such a nonvolatile memory based on a quantum geometric property with non-destructive electrical reading, appealing for neuromorphic computing paradigms[35,36]. Finally, the ability to control Berry curvature and crystal symmetry via stacking transitions may enable the exploration of the uncharted interplay between these degrees of freedom and recently proposed higher order topologies[37–39].

## Methods

### Device fabrication

The dual-gated few-layer $WTe_2$ devices were fabricated in the following sequence: First, graphite and h-BN crystals were mechanically exfoliated onto 280 nm $SiO_2$/Si substrates. Graphite flakes 2–5 nm thick were chosen for the top and bottom gates and 10 - 30-nm-thick h-BN flakes were chosen for the top and bottom dielectric. The top and bottom parts were prepared separately using a polymer-based dry transfer technique. For the bottom part, an h-BN flake was picked up on a polymer stamp and placed on the bottom graphite. After dissolving the polymer, fine Pt metal contacts (~ 5 nm) were patterned on the h-BN. The surface was further cleaned by a 400 °C annealing process in high vacuum. For the top part, the graphite was picked up by another polymer stamp first, then the top h-BN. Both stacks were then transferred to an oxygen- and water-free glovebox. $WTe_2$ crystals were exfoliated inside the glovebox and flakes with suitable thickness were identified by optical contrast. Then such flakes were picked up with the top part; the full stack was then completed by transferring the top stack (graphite/BN/$WTe_2$) onto contacts/h-BN/graphite bottom stack before taking out of the glovebox. Finally, after dissolving the polymer, another step of e-beam lithography and evaporation was used to define electrical bonding pads (Cr/Au) connecting to the metal contacts and the top and bottom gates. The layer number of $WTe_2$ and thickness of BN were confirmed by Raman spectroscopy and AFM[29].

### SHG, Raman spectroscopy and coherent phonon pump-probe dynamics

SHG spectroscopy: The excitation light centered at 800 nm was extracted from a mode-locked 80 MHz titanium-sapphire oscillator. The laser light was focused with a 40X long working distance objective on the sample located in a continuous-flow liquid-nitrogen cryostat. The SHG signal was detected in a backscattering configuration and finally collected by a single-photon PMT counter from Hamamatsu with suitable bandpass filters. In all gate dependent SHG measurements, the excitation light polarization is along *b* axis without analyzer. The excitation light incidence angle is at either 0 degree (normal incidence) or 30 degree (oblique incidence). In SHG polarization pattern measurements, *s* polarization is along the *a* axis while *p* polarization is along the *b* axis. All SHG measurements were conducted at 80 K unless explicitly noted.

Raman spectroscopy was performed using a commercial Raman system (Horiba Labram HR Evolution) with a helium-neon laser ($\lambda = 632.8$ nm) at normal incidence. The system is installed with an ultralow-frequency module to allow detection of ultralow Raman scattering down to 7 $cm^{-1}$. The laser beam was focused with a spot size ~ 1.5 μm on the samples by a 40 X long working distance objective with a correction ring (N.A. = 0.6); The samples were located in a continuous-flow liquid-nitrogen cryostat which can be cooled down to 78 K in a vacuum of $10^{-5}$ mbar. A polarized Raman measurement was achieved by using a visible half-wave plate mounted on a motorized stage before the objective. The incident polarization of the linear polarized laser can be accurately controlled by rotating the half-wave plate. The Raman signal was collected in a reflection configuration without an extra polarizer and plotted after the subtraction of reference background. All Raman measurements were conducted at 80 K unless explicitly noted.

A non-degenerate pump-probe setup was used to study the coherent phonon dynamics in few-layer $WTe_2$ devices. The pump beam (~ 1300 nm) is from an optical parametric oscillator (OPO) pumped by a mode-locked titanium-sapphire oscillator while the probe beam (~ 800 nm) is from the same titanium-sapphire oscillator. The pump beam was modulated by a chopper at a frequency of 2 kHz and focused onto the sample to excite coherent phonons, while the reflectivity change of the probe beam was measured by a silicon-based photodetector coupled to a SR830 lock-in amplifier. The excited coherent phonons can modulate the optical susceptibility at the phonon frequency, which result in corresponding reflectivity oscillations of the probe beam. We keep the pump fluence at about 300 μJ/$cm^2$, 10 times stronger than the probe pulses.

**Gate dependent nonlinear Hall measurements**

Electrical transport was measured in a continuous-flow liquid-nitrogen cryostat. The top and bottom gate voltages were applied through two Keithley 2450 sourcemeters. First- and second-harmonic signals were collected simultaneously by standard lock-in techniques by two SR830 lock-in amplifiers with excitation frequency at 50 Hz. The phase of the first-harmonic (second-harmonic) signal was approximately 0° (±90°), consistent with the expected values for first- and second-order responses. All electrical measurements were conducted at 80 K unless explicitly noted.

**First-principles electronic structure calculations**

First-principles density functional theory was applied for structural relaxation and electric polarization calculation using Vienna Ab initio Simulation Package (VASP) with the Perdew-Burke-Ernzerhof exchange-correlation functional[40,41], a plane-wave basis with cutoff of 300 eV, a 6×12×1 Monkhorst-Pack k-point sampling, and optB88-vdW functional for interlayer van der Waals interactions[42]. Furthermore, to compute Berry curvature, we developed first-principles tight-binding Hamiltonian in the quasiatomic Wannier function basis set based on the maximal similarity measure[43,44]. Here we include spin-orbit coupling and adopt HSE06 hybrid exchange-correlation energy functional with the range-separation parameter $\lambda$ = 0.2[45]. Using the above approach, we then developed effective tight-binding Hamiltonian for trilayer and four-layer $WTe_2$ with total 168 and 224 quasiatomic Wannier functions, respectively. Berry curvature was subsequently calculated with a dense k-point sampling of 300×300×1.

## Data availability

Source data are available for this paper. All other data that support the plots within this paper and other findings of this study are available from the corresponding author upon reasonable request.

## References in Methods Section:

**Acknowledgments:**

This work is supported by the US Department of Energy (DOE), Office of Basic Energy Sciences, Division of Materials Sciences and Engineering, under contract number DE-AC02-76SF00515 (J.X., E.J.S., C.M.N., P.M., C.D.P., T.P.D., A.M.L.). E.J.S. acknowledges additional support from Stanford GLAM Postdoctoral Fellowship Program. C.M.N. acknowledges additional support from the National Science Foundation (NSF) through a Graduate Research Fellowship (DGE-114747). H.W. and X.Q. acknowledge the support by the National Science Foundation (NSF) under award number DMR-1753054. J.X., A.M.L., and C.D.P. acknowledge support for theory calculations through the Center for Non-Perturbative Studies of Functional Materials. Y. W., S.W., and X. Z. acknowledge the support from the U.S. Department of Energy, Office of Science, Office of Basic Energy Sciences, Materials Sciences and Engineering Division within the van der Waals Heterostructures Program (KCWF16) under contract No. DEAC02-05-CH11231 for electrical transport measurement, and the support from King Abdullah University of Science and Technology (KAUST) Office of Sponsored Research award OSR-2016-CRG5-2996 for device design and fabrication. First-principles electronic structure and Berry curvature calculations by H.W. and X.Q. were conducted with the advanced computing resources provided by Texas A&M High Performance Research Computing. Part of this work was performed at the Stanford Nano Shared Facilities (SNSF)/Stanford Nanofabrication Facility (SNF), supported by the National Science Foundation under award ECCS-1542152.

## Author contributions

A.M.L. and X.Z. supervised the project; J.X. and A.M.L. conceived the research; J.X. and Y.W. performed the optical and electrical experiments; Y.W., J.X., S.W., P. M. fabricated the devices; H.W. and X.Q. performed first-principles calculations on the band structure and the Berry curvature through the stacking transitions; C.D.P. conducted theoretical calculations on crystal structures under the supervision of T.P.D.; J.X., Y.W., E.J.S., C.M.N., S.W., P.M. analyzed and interpreted the data with A.M.L. and X.Z.; All authors contributed to the writing of the manuscript.

## Competing interests

J.X. and A.M.L. have submitted a patent application ("Low-energy cost Berry curvature memory based on nanometer-thick layered materials"; US number 62/940,181) that covers a specific aspect of the manuscript. The other authors declare no competing interests.

## Main Figures

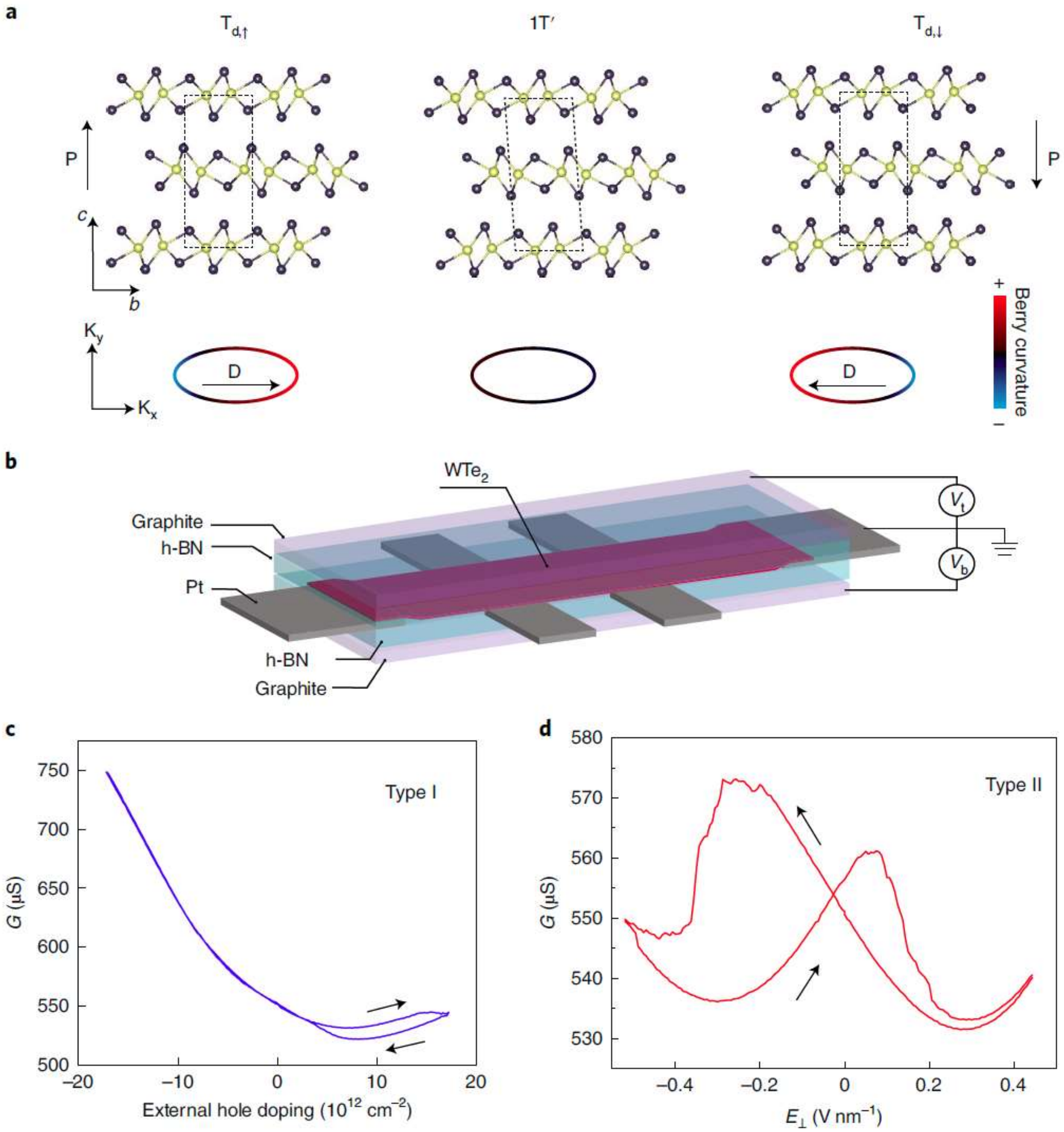

**Figure 1| Signatures of two different electrically-driven phase transitions in $WTe_2$. a,** Side view (*b–c* plane) of unit cell showing possible stacking orders in $WTe_2$ (monoclinic 1T', polar orthorhombic $T_{d,\uparrow}$ or $T_{d,\downarrow}$) and schematics of their Berry curvature distributions in momentum space. The spontaneous polarization and the Berry curvature dipole are labelled as P and D, respectively. The yellow spheres refer to W atoms while the black spheres represent Te atoms. **b,** Schematic of dual-gate h-BN capped $WTe_2$

device. **c,** Electrical conductance G with rectangular-shape hysteresis (labeled as Type I) induced by external doping at 80 K. Pure doping was applied following $V_t/d_t = V_b/d_b$ under a scan sequence indicated by black arrows. **d,** Electrical conductance G with butterfly-shape switching (labeled as Type II) driven by electric field at 80 K. Pure E field was applied following $-V_t/d_t = V_b/d_b$ under a scan sequence indicated by black arrows. Positive $E_\perp$ is defined along +*c* axis. Based on the distinct hysteresis observations in **c** and **d**, two different phase transitions can be induced by different gating configurations.

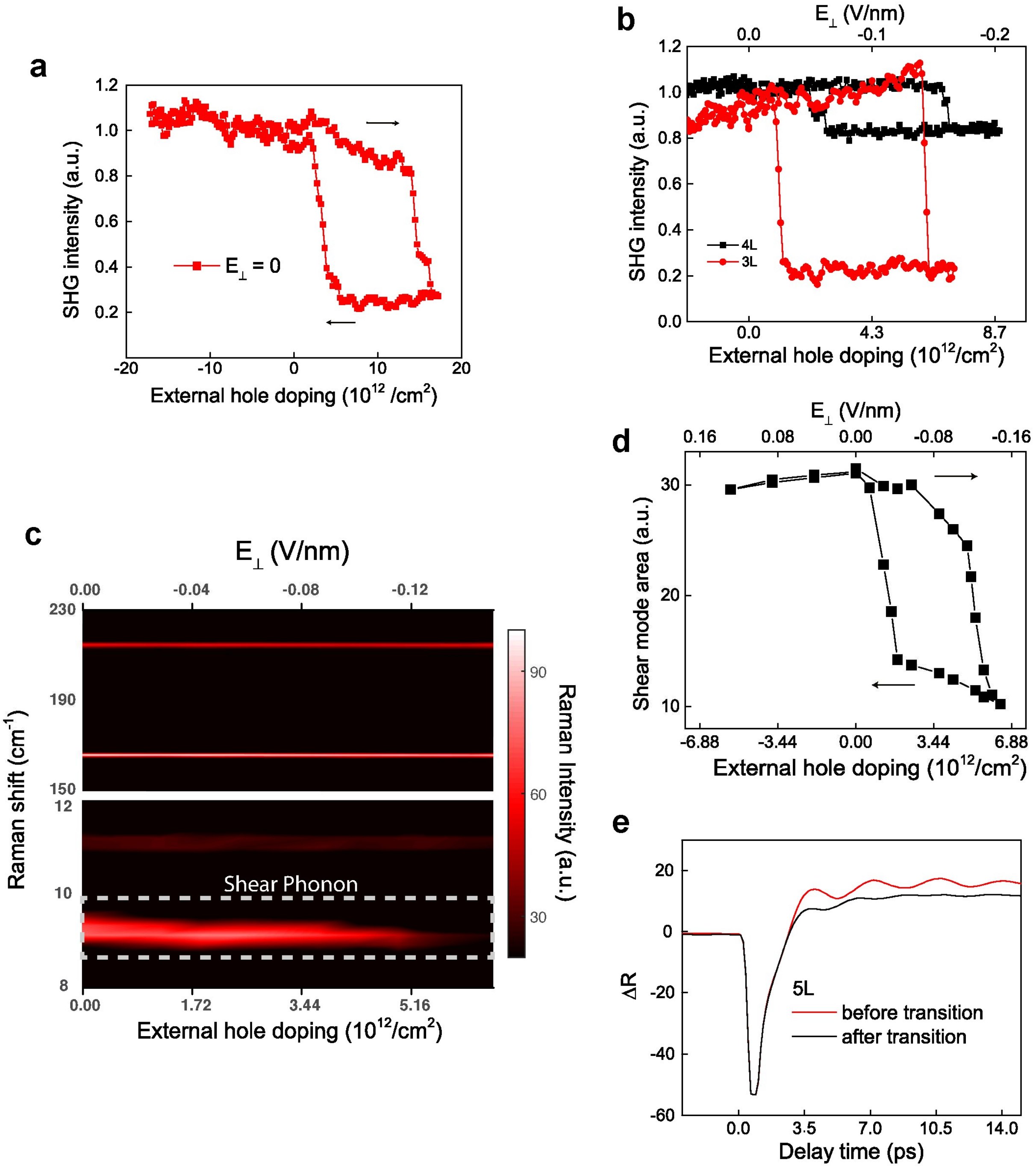

**Figure 2| Observation of transition between $T_d$ and 1T' stackings as the origin for Type I hysteresis.** **a,** A rectangular-shape SHG hysteresis with deep SHG modulation depth under pure doping sweep is observed (black arrows indicate the sweeping direction), correlated with Type I conductance hysteresis in fig 1c. The hole doping side is more favorable to drive the transition. **b,** Layer dependent modulation depth

in rectangular-shape SHG hysteresis. Here we normalize the initial intensity and focus on the relative intensity variation when sweeping the electrical bias. The SHG drop is much larger in odd-layer compared with that in even-layer crystals. Such SHG contrast of the electrically induced state is in line with the layer dependent inversion symmetry preservation and breaking for few-layer 1T' stacking. **c,** Raman spectra evolution of intralayer and interlayer vibrations during Type I phase transition in five-layer $WTe_2$. Intralayer vibrations (165 and 215 $cm^{-1}$) display negligible modulation in intensity and frequency. In contrast, substantial intensity reduction of the *b* axis interlayer shear mode (9 $cm^{-1}$) was observed, consistent with the formation of centrosymmetric 1T' stacking in this five-layer sample. The intensities of the intralayer vibrations have been divided by 10 to compare with the interlayer vibrations within the same color map plot. **d,** Integrated area of such interlayer shear mode shows hysteresis response corresponding to the Type I transition. **e,** Coherent shear phonon dynamics in a five-layer device shows disappearance of coherent shear phonon vibrations after the transition. $\Delta R$ is the transient reflectivity change in arbitrary units. Together this shows that the origin of Type I phase change is a stacking transition between $T_d$ and 1T' through interlayer sliding along the crystalline *b* axis, excluding any intralayer bond distortion or bond breaking.

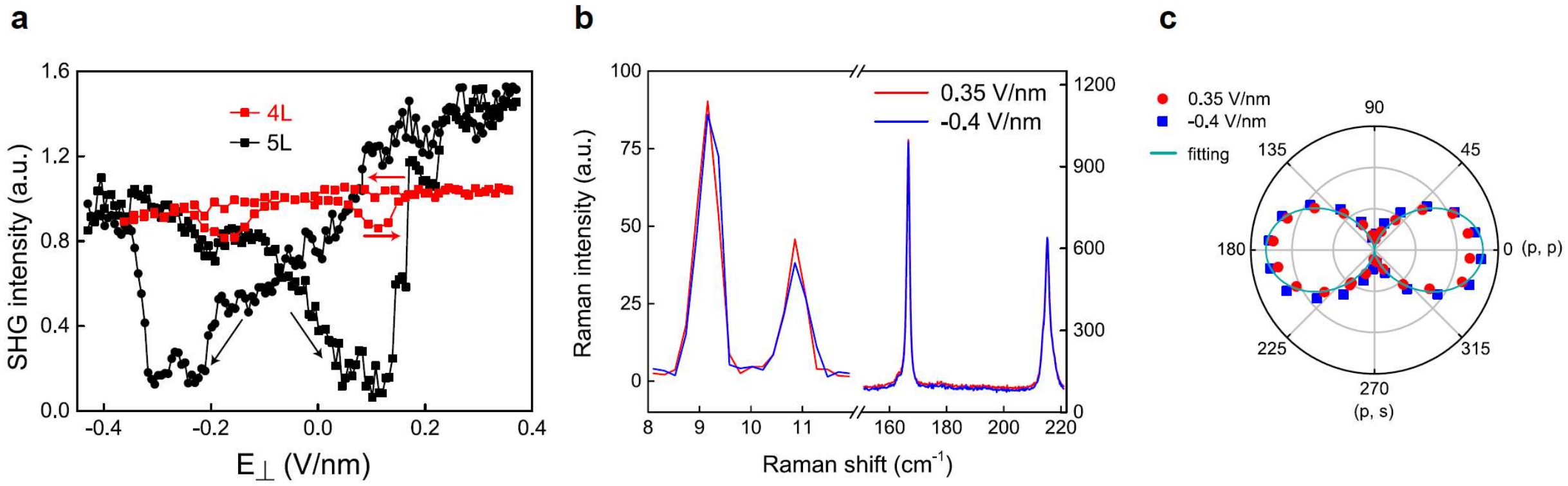

**Figure 3| $T_{d,}$↑ to $T_{d,}$↓ stacking transitions with preserved crystal orientation in Type II hysteresis. a,** *in-situ* SHG intensity evolution in Type II phase transition, driven by a pure E field sweep on a four-layer and a five-layer $T_d$-$WTe_2$ devices (indicated by the arrows). Both show butterfly-shape SHG intensity hysteresis responses as a signature of ferroelectric switching between upward and downward polarization phases. The intensity minima at turning points in four-layer and five-layer crystals show significant difference in magnitude, consistent with the layer dependent SHG contrast in 1T' stacking. This suggests changes in stacking structures take place during the Type II phase transition, which may involve 1T' stacking as the intermediate state. **b,** Raman spectra of both interlayer and intralayer vibrations of fully poled upward and downward polarization phases in the 5L sample, showing nearly identical characteristic phonons of polar $T_d$ crystals. **c,** SHG intensity of fully poled upward and downward polarization phases as a function of analyzer polarization angle, with fixed incident polarization along *p* direction (or *b* axis). Both the polarization patterns and lobe orientations of these two phases are almost the same and can be well fitted based on the second order susceptibility matrix of P*m* space group (Supplementary Information Section I). These observations reveal the transition between $T_{d,}$ ↑ and $T_{d,}$ ↓ stacking orders is the origin of Type II phase transition, through which the crystal orientations are preserved.

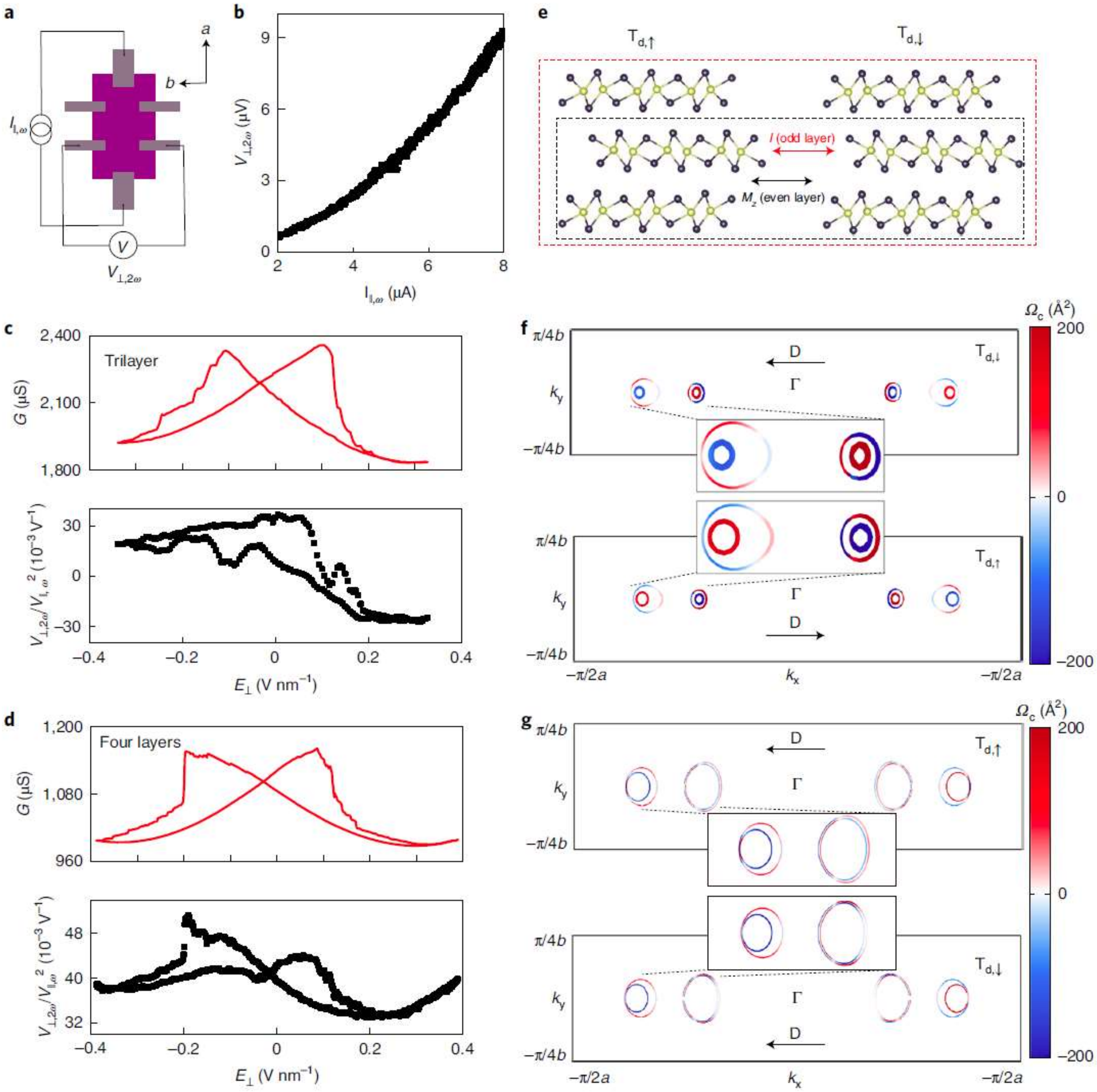

**Fig. 4| Layer-parity selective Berry curvature memory behavior in $T_{d,\uparrow}$ to $T_{d,\downarrow}$ stacking transition. a,** The nonlinear Hall effect measurement schematics. An applied current flow along the *a* axis results in the generation of nonlinear Hall voltage along the *b* axis, proportional to the Berry curvature dipole strength at the Fermi level. **b,** Quadratic amplitude of nonlinear transverse voltage at 2ω as a function of longitudinal current at ω. **c**, **d,** Electric field dependent longitudinal conductance (upper figure) and nonlinear Hall signal (lower figure) in trilayer $WTe_2$ and four-layer $WTe_2$ respectively. Though similar butterfly-shape hysteresis in longitudinal conductance are observed, the sign of the nonlinear Hall signal was observed to be reversed

in the trilayer while maintaining unchanged in the four-layer crystal. Because the nonlinear Hall signal ($V_{\perp,2\omega}$ / $(V_{//,\omega})^2$ ) is proportional to Berry curvature dipole strength, it indicates the flipping of Berry curvature dipole only occurs in trilayer. **e,** Schematics of layer-parity selective symmetry operations effectively transforming $T_{d,}\uparrow$ to $T_{d,}\downarrow$. The interlayer sliding transition between these two ferroelectric stackings is equivalent to an inversion operation in odd layer while a mirror operation respect to the *ab* plane in even layer. **f, g,** Calculated Berry curvature $\Omega_c$ distribution in 2D Brillouin zone at the Fermi level for $T_{d,}\uparrow$ and $T_{d,}\downarrow$ in trilayer and four-layer $WTe_2$. The symmetry operation analysis and first principle calculations confirm Berry curvature and its dipole sign reversal in trilayer while invariant in four-layer, leading to the observed layer-parity selective nonlinear Hall memory behavior.